\documentclass[12pt]{article}

\usepackage{amsmath}
\numberwithin{equation}{section}

\evensidemargin \oddsidemargin

\begin{document}

\begin{titlepage}

\renewcommand{\thefootnote}{\fnsymbol{footnote}}

\hfill{hep-th/0212343}

\vspace{15mm}
\baselineskip 9mm
\begin{center}
{\LARGE \bf Covariant Description of D-branes\\
in IIA Plane-Wave Background}
\end{center}

\baselineskip 6mm
\vspace{10mm}
\begin{center}
Seungjoon Hyun,$^a$\footnote{\tt hyun@phya.yonsei.ac.kr} 
Jaemo Park$^b$\footnote{\tt jaemo@physics.postech.ac.kr} 
and Hyeonjoon Shin$^c$\footnote{\tt hshin@newton.skku.ac.kr} 
\\[5mm] 
{\sl $^a$Institute of Physics and Applied Physics, 
Yonsei University, Seoul 120-749, Korea \\ 
$^b$ Department of Physics, POSTECH, Pohang 790-784,
Korea \\ 
$^c$ BK 21 Physics Research Division and Institute of Basic
Science \\ Sungkyunkwan University, Suwon 440-746, Korea}
\end{center}

\vspace{20mm}
\begin{center}
{\sc Dedicated to the memory of Youngjai Kiem}
\end{center}
\vspace{20mm}
\thispagestyle{empty}

\vfill
\begin{center}
{\bf Abstract}
\end{center}
\noindent
We work out boundary conditions for the covariant open string in the
type IIA plane wave background, which corresponds to the D-branes in
the type IIA theory. We use the kappa symmetric string action and see
what kind of boundary conditions should be imposed to retain kappa
symmetry.  We find half BPS as well as quarter BPS branes and the
analysis agrees with the previous work in the light cone gauge if the
result is available.  Finally we find that D0-brane is
non-supersymmetric.
 
\vspace{20mm}
\end{titlepage}

\baselineskip 6.5mm
\renewcommand{\thefootnote}{\arabic{footnote}}
\setcounter{footnote}{0}

\section{Introduction} 

Recently the string theory on the plane wave background has attracted
much attention in relation to the correspondence to the N=4
Supersymmetric Yang-Mills (SYM) theory \cite{met044,ber021}. It is now
well known that for the usual AdS/CFT correspondence, the various
checks have been done essentially on the supergravity states on the
string side.  In the seminal paper by Berenstein, Maldacena and
Nastase \cite{ber021}, this major obstacle was overcome, thereby
showing the explicit correspondence between more general string states
and the suitable Yang-Mills operators in the plane wave background
\cite{bla242}, which is the Penrose limit of the AdS \cite{bla081}.
After their paper, more progress was made on how the string
Hamiltonian is mapped to the anomalous dimension of the Yang-Mills
operator and more precise dictionaries for the correspondence have
been developed \cite{sug, sug1, parity, str, ym, mix, bit}.

The previously mentioned development was made in Type IIB side.  In
Type IIA side, the matrix theory on the plane wave background has been
important focus in relation to the better understanding of the
M-theory on the plane wave background \cite{ber021,das185}.  Recently,
simple Type IIA string theory on the plane wave background was
proposed by the Kaluza-Klein compactification of the M-theory
\cite{hyu074,sug029}.  The resulting string theory has many nice
features. It admits light cone gauge where the string theory spectrum
is that of the free massive theory as happens in Type IIB
\cite{met044}.  Furthermore the worldsheet enjoys (4,4) supersymmetry
\cite{hyu074}.  The structure of the supersymmetry is simpler than
that of Type IIB in the sense that the supersymmetry commutes with the
Hamiltonian so that all members of the same supermultiplet has the
same mass.  In the extended version, more detailed exposition was
given and the various 1/2 BPS D-branes states were analyzed in the
light cone gauge \cite{hyu158}, which are compatible with the BPS
branes in matrix model \cite{hyu090,par161}.

The purpose of this letter is to carry out the analysis about the
D-branes states in the covariant setting. We follow the logic of
Lambert and West \cite{lam031} and consider the kappa symmetric string
action in the plane wave background and figure out how the
supersymmetry is reduced when we impose suitable boundary conditions
on the boundary of the worldsheet.  Similar work \cite{bai038} has
been done in Type IIB side and given results in agreement with the
previous results \cite{dab231,bil028,ske054}.  See also \cite{gab122,dpp}
for recent alternative study. Various aspects of kappa symmetry 
and worldsheet supersymmetry in the plane wave background is discussed
in \cite{cvetic}. In our study, as expected, for the
D-brane located at the origin of the plane wave background, the
analysis in the covariant setting coincides with that in the lightcone
gauge \cite{hyu158}. The merit of the covariant analysis is that we
are able to work out other D-brane states, which are difficult to
analyze in the lightcone gauge.  For example, we work out the
supersymmetry of D-particle (in fact, nonsupersymmetry) and analyze
the supersymmetry of D-branes located away from the origin.  In the
investigation we found out some potential subtleties arising in the
D-brane analysis in the lightcone gauge.  We will comment on this in
the later section.  We think that the analysis in the covariant
setting is a good starting point to sort out various, sometimes
conflicting, claims \cite{ske184} on the number of supersymmetry of
various D-branes in the PP-wave.

\section{Covariant Wess-Zumino action of Type IIA string}

The covariant description of D-branes via open string may be given by
investigating the boundary contributions in the kappa symmetry
variation of the Wess-Zumino part of the superstring action
\cite{bai038}.  In this section, starting from the superspace geometry
of $AdS_4 \times S^7$ \cite{dew209} whose Penrose limit leads to the
eleven dimensional pp-wave background, we derive the covariant
Wess-Zumino action of Type IIA superstring in the IIA pp-wave
background of Refs. \cite{hyu074,sug029} up to quartic order in the
fermionic coordinate $\theta$.

The eleven dimensional superspace geometry of $AdS_4 \times S^7$
\cite{dew209} is encoded in the super elfbein $\widehat{E}^{\hat{A}} =
(\widehat{E}^{\hat{r}}, \widehat{E})$ and the three form superfield
$\widehat{B}$.\footnote{We note that $\widehat{E}$ means
  $\widehat{E}^{\hat{a}}$.  The index notations adopted here are as
  follows: $M,N,...$ are used for the target superspace indices while
  $A,B,...$ for tangent superspace. As usual, a superspace index is
  the composition of two types of indices such as $M=(\mu,\alpha)$ and
  $A=(r,a)$. $\mu,\nu,...~(r,s,...)$ are the ten dimensional target
  (tangent) space-time indices.  $\alpha,\beta,...~(a,b,...)$ are the
  ten dimensional (tangent) spinor indices. For the eleven dimensional
  case, we denote quantities and indices with hat to distinguish from
  those of ten dimensions.  $m,n,...$ are the worldsheet vector
  indices with values $\tau$ and $\sigma$. The convention for the
  worldsheet antisymmetric tensor is taken to be
  $\epsilon^{\tau\sigma}=1$. }  The super elfbein is
\begin{eqnarray}
\widehat{E}^{\hat{r}}
        &=& dx^{\hat{\mu}} \hat{e}^{\hat{r}}_{\hat{\mu}}
          + 2 \sum^{15}_{n=0} \frac{1}{(2n+2)!}
           \bar{\theta} \Gamma^{\hat{r}} {\cal M}^{2n} 
           \widehat{D} \theta ~,
     \nonumber \\
\widehat{E} &=& \sum^{16}_{n=0} \frac{1}{(2n+1)!} {\cal M}^{2n} 
             \widehat{D} \theta ~,
\label{self}
\end{eqnarray}
where $\hat{e}^{\hat{r}}_{\hat{\mu}}$ is the elfbein and $\widehat{D}
\theta$, the covariant derivative of $\theta$, is given by
\begin{equation}
\widehat{D} \theta
= d \theta 
 + \frac{1}{4} \hat{\omega}^{\hat{r}\hat{s}}
   \Gamma_{\hat{r}\hat{s}}
 + \hat{e}^{\hat{r}} 
   T_{\hat{r}}{}^{\hat{s}\hat{t}\hat{u}\hat{v}}
   \widehat{F}_{\hat{s}\hat{t}\hat{u}\hat{v}} \theta
\end{equation}
with the eleven dimensional spin connection
$\hat{\omega}^{\hat{r}\hat{s}}$. The matrix ${\cal M}^2$ is
\begin{equation}
{\cal M}^2_{ab} 
= 2 (T_{\hat{r}}{}^{\hat{s}\hat{t}\hat{u}\hat{v}}
     \widehat{F}_{\hat{s}\hat{t}\hat{u}\hat{v}} \theta)_a 
    (\bar{\theta} \Gamma^{\hat{r}})_b
-\frac{1}{4} 
    (\Gamma^{\hat{r}\hat{s}} \theta)_a
    (\bar{\theta} S_{\hat{r}\hat{s}}{}^{\hat{t}\hat{u}\hat{v}\hat{w}}
     \widehat{F}_{\hat{t}\hat{u}\hat{v}\hat{w}})_b ~.
\end{equation}
The definitions for the tensor structures are as follows:
\begin{equation}
T_{\hat{r}}{}^{\hat{s}\hat{t}\hat{u}\hat{v}}
\equiv \frac{1}{288} 
  (\Gamma_{\hat{r}}{}^{\hat{s}\hat{t}\hat{u}\hat{v}}
  - 8 \delta_{\hat{r}}^{ [ \hat{s}} 
      \Gamma^{\hat{t}\hat{u}\hat{v}]}
  ) ~,~~~
S_{\hat{r}\hat{s}}{}^{\hat{t}\hat{u}\hat{v}\hat{w}}
\equiv \frac{1}{72}
  (\Gamma_{\hat{r}\hat{s}}{}^{\hat{t}\hat{u}\hat{v}\hat{w}}
  + 24 \delta_{\hat{r}}^{ [ \hat{t}} \delta_{\hat{s}}^{\hat{u}}
    \Gamma^{\hat{v}\hat{w}]} 
  ) ~.
\end{equation}
The three form superfield is given by
\begin{equation}
\widehat{B} 
= \frac{1}{6} \hat{e}^{\hat{r}} \wedge \hat{e}^{\hat{s}} \wedge
    \hat{e}^{\hat{t}} \widehat{C}_{\hat{r}\hat{s}\hat{t}}
 - \int^1_0 dt \; \bar{\theta} \Gamma_{\hat{r}\hat{s}}
    \widehat{E} \wedge \widehat{E}^{\hat{r}} 
    \wedge \widehat{E}^{\hat{s}} ~,
\label{s3f}
\end{equation}
where $\widehat{C}_{\hat{r}\hat{s}\hat{t}}$ is three form gauge field
whose field strength is $\widehat{F}_{\hat{r}\hat{s}\hat{t}\hat{u}} =
4 \partial_{[ \hat{r}} \widehat{C}_{\hat{s}\hat{t}\hat{u} ] }$.  We
note that the super elfbeins in the second term on the right hand side
have $t$ dependence in a way that $\theta$'s in (\ref{self}) are
replaced as $\theta \rightarrow t \theta$.

The component fields in Eqs. (\ref{self}) and (\ref{s3f}) are for the
$AdS_4 \times S^7$.  As shown in \cite{bla081}, by taking the Penrose
limit \cite{pen271}, they become the fields describing the eleven
dimensional pp-wave background.  After some rotation in a certain
plane, say 49-plane, for our convenience in ten dimensions, the eleven
dimensional pp-wave background becomes as follows:\footnote{For
detailed derivation of this background and its ten dimensional
reduction, see \cite{hyu074}.}
\begin{eqnarray}
& & \hat{e}^+ = dx^+ ~,~~~
    \hat{e}^- = dx^- + \frac{1}{2} A(x^I) dx^+ ~,
  \nonumber \\
& & \hat{e}^I = dx^I ~,~~~
    \hat{e}^9 = dx^9 + \frac{\mu}{3} x^4 dx^+ ~,
  \nonumber \\
& & \widehat{F}_{+123} = \mu ~,
\label{11pp}
\end{eqnarray}
where $\mu$ is constant characterizing the pp-wave, $I = 1,...,8$ and
\begin{equation}
A(x^I) = \left( \frac{\mu}{3} \right)^2 
          \sum^4_{i=1} (x^i)^2
         +\left( \frac{\mu}{6} \right)^2
          \sum^8_{i'=5} (x^{i'})^2~.
\label{pot}
\end{equation}

We now turn to the ten dimensional background, which is the Type IIA
pp-wave background obtained from the circle compactification of the
eleven dimensional pp-wave (\ref{11pp}).  If we take $x^9$ as the
direction of compactification, then the usual Kaluza-Klein dimensional
reduction leads us to have the following ten dimensional background:
\begin{eqnarray}
 & & e^+ = dx^+ ~,
   \nonumber \\
 & & e^- = dx^- + \frac{1}{2} A(x^I) dx^+ ~,
   \nonumber \\
 & & e^I = dx^I ~,
   \nonumber \\
 & & F_{+123} = \mu~,~~~ F_{+4} = - \frac{\mu}{3} ~.
\label{pp}
\end{eqnarray}

In terms of these ten dimensional fields and by using the logic of the
Kaluza-Klein reduction, one can express the elementary pieces of the
eleven dimensional superfields (\ref{self}) and (\ref{s3f}), which are
$\widehat{D} \theta$ and the matrix ${\cal M}^2$.  Firstly, the eleven
dimensional supercovariant derivative becomes
\begin{equation}
\widehat{D} \theta = 
  D \theta + \frac{\mu}{6} (\Gamma^{+4} h_- \theta) \hat{e}^9 ~,
\label{11covd}
\end{equation}
where $D \theta$ is the ten dimensional supercovariant derivative of
$\theta$ and $h_\pm$ is the operator projecting spinor states onto the
states with eigenvalue $\pm 1$ of $\Gamma^{12349}$;
\begin{equation}
h_\pm = \frac{1}{2} ( 1 \pm \Gamma^{12349} ) ~.
\end{equation}
The covariant derivative $D \theta$ is given by
\begin{equation}
D \theta =
     d \theta + \frac{1}{4} \omega^{rs} \Gamma_{rs} \theta
     + \Omega \theta ~,
\end{equation}
where the non-vanishing ten dimensional spin connection is
\begin{equation}
\omega^{-I} = \frac{1}{2} \partial_I A dx^+
\end{equation}
and the definition for $\Omega$ is 
\begin{eqnarray}
\Omega
 &=& \frac{\mu}{12} 
    \Big[ - e^+ \left( \Gamma^- \Gamma^{+123} 
              + 2 \Gamma^{49} ( 2 h_- - h_+ )
            \right)
          + 2 e^i \Gamma^{+i} \Gamma^{123}
          + 2 e^4 \Gamma^{+9} h_+
          - e^{i'} \Gamma^{+ i'} \Gamma^{123}  
   \Big] ~,
\nonumber \\
\end{eqnarray}
where $i=1,2,3$ and $i' = 5,6,7,8$.  The explicit expression of matrix
${\cal M}^2$ in term of ten dimensional quantities is obtained as
\begin{eqnarray}
{\cal M}^2_{ab}
 &=& - i \frac{\mu}{6} 
  \bigg[ \, 
       ( \Gamma^- \Gamma^{+123} \theta + 3 \Gamma^{123} \theta)_a
       (\bar{\theta} \Gamma^+ )_b
     + (\Gamma^{+ i'} \Gamma^{123} \theta)_a
       (\bar{\theta} \Gamma^{i'} )_b
     + (\Gamma^{+4} \Gamma^{123} \theta)_a
       (\bar{\theta} \Gamma^4 )_b
\nonumber \\
 & & \hspace{10mm}
     + (\Gamma^{+9} \Gamma^{123} \theta)_a
       (\bar{\theta} \Gamma^9 )_b
     + (\Gamma^{+ i'} \theta)_a
       (\bar{\theta} \Gamma^{-+ i'} \Gamma^{123} )_b
     + (\Gamma^{+ 4} \theta)_a
       (\bar{\theta} \Gamma^{-+ 4} \Gamma^{123} )_b
\nonumber \\
 & & \hspace{10mm}
     + (\Gamma^{+ 9} \theta)_a
       (\bar{\theta} \Gamma^{-+ 9} \Gamma^{123} )_b
     +\frac{1}{2} (\Gamma^{i' j'} \theta)_a
       (\bar{\theta} \Gamma^{+ i' j'}\Gamma^{123})_b
     + (\Gamma^{ i' 4} \theta)_a
           (\bar{\theta} \Gamma^{+ i' 4}\Gamma^{123})_b
\nonumber \\
 & & \hspace{10mm}
     + (\Gamma^{ i' 9} \theta)_a
       (\bar{\theta} \Gamma^{+ i' 9}\Gamma^{123})_b
     + (\Gamma^{+ i j} \theta)_a
       (\bar{\theta} \Gamma^{ij} \Gamma^{123})_b
     - (\Gamma^{+ij} \Gamma^{123} \theta)_a
       (\bar{\theta} \Gamma^{ij})_b
\nonumber \\
 & & \hspace{10mm}
     - (\Gamma^{ij} \theta)_a
       (\bar{\theta} \Gamma^{+ij} \Gamma^{123})_b
  \bigg] ~.
\label{matm}
\end{eqnarray}
Here we note that, going from eleven to ten dimensions, the fermionic
coordinate $\theta$ splits into two Majorana-Weyl spinors with
opposite $SO(1,9)$ chiralities measured by $\Gamma^9$:
\begin{equation}
\theta = \theta^1 + \theta^2 ~,
\end{equation}
where $\Gamma^9 \theta^1 = \theta^1$ and $\Gamma^9 \theta^2 = -
\theta^2$.

We now have all the ingredients for writing down the covariant
Wess-Zumino action of Type IIA string in the pp-wave background,
(\ref{pp}).  In the superfield formalism, the Wess-Zumino action
is given by
\begin{equation}
S_{WZ} 
= \frac{1}{2 \pi \alpha'} \int_\Sigma d^2 \sigma
     \frac{1}{2!} \epsilon^{mn} \Pi^A_m \Pi^B_n B_{BA} ~,
\end{equation}
where $\Pi^A_m = \partial_m Z^M E^A_M$ with supercoordinate $Z^M =
(X^\mu, \theta^\alpha)$ and $B_{BA}$ is the two form superfield.
$\Sigma$ represents the worldsheet of open string.  In the context of
this paper, it is useful to remind the well known fact that $S_{WZ}$
can be viewed as the action obtained from the Wess-Zumino action for
the eleven dimensional super membrane through the double dimensional
reduction \cite{duf70}.  Since we compactify the super membrane along
the $x^9$ direction, the two form superfield is identified with the
eleven dimensional three form superfield with the index $9$, that is,
$\widehat{B}_{9NM}$. Then the above Wess-Zumino action is rewritten as
\begin{equation}
S_{WZ} = \frac{1}{2 \pi \alpha'} \int_\Sigma d^2 \sigma
     \frac{1}{2!} \epsilon^{mn} \partial_m Z^M \partial_n Z^N
       \widehat{B}_{9NM}
\label{swz} ~.
\end{equation} 
Now by using Eqs. (\ref{self}), (\ref{s3f}), (\ref{pp}),
(\ref{11covd}), and (\ref{matm}), the Wess-Zumino action is given
in component form and expanded in terms of $\theta$.  Although it
has expansion up to 32th order in $\theta$, we will give the expansion
up to quartic order since, as we shall see in the next section, 
the nontrivial information for the description of D-branes is obtained
already at the quartic order.  The resulting Wess-Zumino action of
Type IIA string in the IIA pp-wave background is then  
\begin{eqnarray}
S_{WZ}
 &=& \frac{i}{2 \pi \alpha'} \int_\Sigma d^2 \sigma \epsilon^{mn}
   \bigg[ 
     - \bar{\theta} \Gamma_{r9} D_m \theta e^r_n
     -\frac{i}{2} (\bar{\theta} \Gamma_{r9} D_m \theta)
                  (\bar{\theta} \Gamma^r D_n \theta)
     -\frac{\mu}{12} 
             (\bar{\theta} \Gamma_{rs} \Gamma^{+4} h_- \theta)
             e^r_m e^s_n
\nonumber \\
 & & + i \frac{\mu}{12} (\bar{\theta} \Gamma^r \Gamma^{+4} h_- \theta)
                 (\bar{\theta} \Gamma_{rs} D_m \theta) e^s_n
     + i \frac{\mu}{12} (\bar{\theta} \Gamma^9 \Gamma^{+4} h_- \theta)
                 (\bar{\theta} \Gamma_{9r} D_m \theta) e^r_n
\nonumber \\
 & & - i \frac{\mu}{12} (\bar{\theta} \Gamma_{rs} \Gamma^{+4} h_- 
\theta)
                 (\bar{\theta} \Gamma^r D_m \theta) e^s_n
     - i \frac{\mu}{12} (\bar{\theta} \Gamma_{9r} \Gamma^{+4} h_- 
\theta)
                 (\bar{\theta} \Gamma^9 D_m \theta) e^r_n
\nonumber \\
 & & - \frac{1}{12} (\bar{\theta} \Gamma_{r9} {\cal M}^2 D_m \theta)
                    e^r_n
     - \frac{\mu}{144} 
        (\bar{\theta} \Gamma_{rs} {\cal M}^2 \Gamma^{+4} h_- \theta)
        e^r_m e^s_n 
     + {\cal O} (\theta^6)
    \bigg] ~,
\label{wz}
\end{eqnarray}
where $e^r_m = \partial_m X^\mu e^r_\mu$ and $D_m \theta = \partial_m
X^\mu (D \theta)_\mu$.

\section{Boundary conditions from kappa symmetry}

The $\kappa$ symmetry transformation rules are read off from 
\begin{equation}
\delta Z^M E_M^r = 0 ~,~~~
\delta Z^M E_M^a = (1-\Gamma \Gamma^9 )^a{}_b \kappa^b  ~,
\label{kappa}
\end{equation}
where $\Gamma = \frac{1}{2!} \epsilon^{mn} \Pi^r_m \Pi^s_n
\Gamma_{rs}$.  From (\ref{kappa}), one can see that the variation of
$X^\mu$ is given by
\begin{equation}
\delta X^\mu
 = - i \bar{\theta}^1 \Gamma^\mu \delta \theta^1
   - i \bar{\theta}^2 \Gamma^\mu \delta \theta^2
   + {\cal O} (\theta^3) ~.
\end{equation}
Here we retain the variations up to the quadratic in $\theta$ since we
are interested in the kappa variation up to the quartic in $\theta$.
As shown in \cite{bai038}, the kinematic parts of the kappa symmetric
action does not produce the boundary terms, so we just consider the
variations of the Wess-Zumino terms. We divide the resulting kappa
variations as three parts, i.e., $\mu$ independent part, $\mu$
dependent part with no position dependence and finally the part with
both $\mu$ and position dependence.

$\mu$ independent part gives the same result as in the flat case,
which gives the well known result \cite{lam031}. The relevant
variation is given by
\begin{eqnarray}
\delta S_{WZ}
 & \rightarrow & \int_{\partial \Sigma}
   \Big[ \; i (\bar{\theta}^1 \Gamma_r \delta \theta^1
           -\bar{\theta}^2 \Gamma_r \delta \theta^2 ) dX^\mu e^r_\mu
 \nonumber \\
 & & \hspace{10mm}
     - ( \bar{\theta}^1 \Gamma_r d \theta^1 
         \bar{\theta}^1 \Gamma^r \delta \theta^1
        -\bar{\theta}^2 \Gamma_r d \theta^2
         \bar{\theta}^2 \Gamma^r \delta \theta^2
       )
    \Big] ~,
\label{b1}
\end{eqnarray}
where $\partial \Sigma$ represents the boundary of $\Sigma$, that is,
the boundary of open string worldsheet.  Here the arrow means that we
are ignoring overall coefficients in front of the Wess-Zumino terms.
In order to have vanishing variation on the boundary, we impose the
usual half-BPS boundary conditions,
\begin{equation}
\theta^2 = P \theta^1
\end{equation}
with
\begin{equation}
P = \Gamma^{+-i_1i_2 \cdots i_{(p-1)}} ~,
\label{P}
\end{equation}
where $+-i_1i_2 \cdots i_{(p-1)}$ denotes the Neumann directions of
the D-brane considered.  And $p$ should be even since $\theta^1$ and
$\theta^2$ have opposite chiralities.  Then\footnote{$r \in N(D)$
  means that $r$ is the direction of Neumann (Dirichlet) boundary
  condition.}
\begin{equation}
\bar{\theta}^2 \Gamma_r \delta \theta^2
 = \left\{  \begin{array}{l@{\quad : \quad}l}
               + \bar{\theta}^1 \Gamma_r \delta \theta^1 &
                r \in N \\
               - \bar{\theta}^1 \Gamma_r \delta \theta^1 &
                r \in D
            \end{array}
   \right. ~.
\end{equation}
It is clear that the boundary conditions eliminate the boundary terms.

The kappa variation of the Wess-Zumino terms which have dependence on
$\mu$ with no position dependence is given by
\begin{eqnarray}
\delta S_{WZ} 
 &\rightarrow&
   \int_{\partial \Sigma}
     dX^\mu e^r_\mu 
\nonumber \\
& & \times
    \bigg[
           ( \bar{\theta}^1 \Gamma^s \delta \theta^1
            +\bar{\theta}^2 \Gamma^s \delta \theta^2
           )
\nonumber \\
 & &   \hspace{10mm}  \times 
     \left( 
       \bar{\theta}^1 \left( 2 \Gamma_{[ r} \Omega_{s]}
            - \frac{\mu}{12} \Gamma_{rs} \Gamma^{+4} h_- \right)
       \theta^2
      -\bar{\theta}^2 \left( 2 \Gamma_{[ r} \Omega_{s]}
            + \frac{\mu}{12} \Gamma_{rs} \Gamma^{+4} h_- \right)
       \theta^1
     \right)
\nonumber \\
 & & +  ( \bar{\theta}^2 \Gamma^s \delta \theta^2 )
        ( \bar{\theta}^1 \Gamma_s \Omega_r \theta^2 ) 
       -( \bar{\theta}^1 \Gamma^s \delta \theta^1 )
        ( \bar{\theta}^2 \Gamma_s \Omega_r \theta^1 ) 
\nonumber \\
 & & - \frac{\mu}{12} 
       (\bar{\theta}^1 \Gamma_{rs} \delta \theta^2
        +\bar{\theta}^2 \Gamma_{rs} \delta \theta^1
       )
       ( \bar{\theta}^1 \Gamma^s \Gamma^{+4} h_- \theta^1
        +\bar{\theta}^2 \Gamma^s \Gamma^{+4} h_- \theta^2
       ) 
\nonumber \\
 & & -\frac{\mu}{12} 
       (\bar{\theta}^1 \delta \theta^2
        -\bar{\theta}^2 \delta \theta^1
       )
       ( \bar{\theta}^1 \Gamma_r \Gamma^{+4} h_- \theta^1
        -\bar{\theta}^2 \Gamma_r \Gamma^{+4} h_- \theta^2
       ) 
\nonumber \\
 & & + \frac{i}{12} 
       (\bar{\theta} \Gamma_{r9} {\cal M}^2 \delta \theta )
  \bigg] ~.
\label{b2}
\end{eqnarray}
First consider terms of the structure $\bar{\theta} \Gamma_{rs} \delta
\theta$.
\begin{equation}
\bar{\theta}^1 \Gamma_{rs} \delta \theta^2
 +\bar{\theta}^2 \Gamma_{rs} \delta \theta^1
= \left\{ \begin{array}{c@{\quad:\quad}l}
              0 & r \in N~,~~ s \in D (N) \quad {\rm for} \quad
                  p = 2,6 ~ (4,8)  \\
          2 \bar{\theta}^1 \Gamma_{rs} P \delta \theta^1 
                & r \in N~,~~ s \in N (D) \quad {\rm for} \quad
                  p = 2,6 ~ (4,8)
          \end{array}
  \right. ~.
\end{equation}
We see that for $p=4,8~(2,6)$ with $s \in D~(N)$, $\bar{\theta}^1
\Gamma^s \Gamma^{+4} h_- \theta^1 +\bar{\theta}^2 \Gamma^s \Gamma^{+4}
h_- \theta^2$ should vanish.  However it does not vanish only with the
boundary condition $\theta^2 = P \theta^1$.  Some constraints should
be imposed on the structure of $P$ and thus picks up the branes with
particular orientations.  Let us label the matrix $P$ by three non
negative integers $n$, $n_4$ and $n'$ with $n+n_4+n'=p-1$:
\[
P^{(n,n_4,n')} ~.
\] 
$n(n')$ denotes the number of gamma matrices with indices in 123
(5678) directions and $n_4$ the presence or the absence of $\Gamma^4$
thus taking value of 0 or 1 respectively in eq. (\ref{P}).

Careful analysis shows that
\begin{equation}
\bar{\theta}^1 \Gamma^s \Gamma^{+4} h_-
\theta^1 +\bar{\theta}^2 \Gamma^s \Gamma^{+4} h_- \theta^2 = 0
\end{equation}
if we bear in mind that $s \in N(D)$ for $p=2,6(4,8)$ and impose the
following constraints,
\begin{eqnarray}
p=2,6 &:& n = {\rm odd}~,~~ n_4 = 0 \nonumber \\
p=4,8 &:& n = {\rm even}~,~~ n_4 = 1 ~.
\label{const}
\end{eqnarray}

Interestingly, one can check, with lengthy calculation, that all other
remaining terms in (\ref{b2}) vanishes if we impose the constraints
(\ref{const}).

Possible D-brane configurations making the above boundary
contributions vanish are given by the following choices of $(n,
n_4,n')$ for $Dp$-brane.
\begin{eqnarray}
p=2 &:& (1,0,0) 
 \nonumber \\
p=4 &:& (0,1,2)~,~~(2,1,0)
 \nonumber \\
p=6 &:& (1,0,4)~,~~(3,0,2) 
 \nonumber \\
p=8 &:& (2,1,4) ~.
\end{eqnarray}
This exactly coincides with the previous result obtained in the
light-cone gauge formulation \cite{hyu158}.

Finally the kappa variations which have both dependence on $\mu$ and
the position are given by
\begin{eqnarray}
\delta S_{WZ}
 & \rightarrow &
  \int_{\partial \Sigma} d X^\mu e^r_\mu
   \bigg[ ( \bar{\theta}^1 \Gamma^s \delta \theta^1
          + \bar{\theta}^2 \Gamma^s \delta \theta^2 )
         ( \bar{\theta}^1 
           \omega_{[r}{}^{-I} \Gamma_{s]} \Gamma^{+I} \theta^1
          -\bar{\theta}^2
           \omega_{[r}{}^{-I} \Gamma_{s]} \Gamma^{+I} \theta^2
         )
\nonumber \\
 & & 
   + \frac{1}{2} (\bar{\theta}^1 \Gamma^s \delta \theta^1)
        (\bar{\theta}^2 \omega_r^{-I} 
          \Gamma_s \Gamma^{+I} \theta^2)
   - \frac{1}{2} (\bar{\theta}^2 \Gamma^s \delta \theta^2)
        (\bar{\theta}^1 \omega_r^{-I} 
          \Gamma_s \Gamma^{+I} \theta^1)
   \bigg] ~.
\label{b3}
\end{eqnarray}
Note that the position dependence comes from the spin connection
$\omega_r^{-I}$. From this action, we need to consider only for $r \in
N$ since $d X^\mu e^r_\mu=0 $ for $ r \in D$. First term vanishes for
$s \in D$ or $I \in N$.  However, for $s \in N$ and $I \in D$, it does
not vanish and becomes
\begin{equation}
4 (\bar{\theta}^1 \Gamma^s \delta \theta^1)
(\bar{\theta}^1 \omega_{[r}{}^{-I} \Gamma_{s]} 
  \Gamma^{+I} \theta^1) ~.
\end{equation}
The remaining terms combine to vanish for $I \in N$ but for $I \in D$
we have
\begin{equation}
\pm (\bar{\theta}^1 \Gamma^s \delta \theta^1)
(\bar{\theta}^1 \omega_r^{-I} \Gamma_s 
  \Gamma^{+I} \theta^1) ~,
\end{equation}
where $+(-)$ sign corresponds to $s \in D (N)$.  At this point, we
have to impose additional boundary condition
\begin{equation}
\Gamma^+ \theta^1 \Big|_{\partial \Sigma} = 0 ~.
\end{equation}
This leads to 1/4-BPS.

What would be the physical consequence of the results obtained above?
It is well known that the appropriate sigma model of the open string
coupled to open string background is given by
\begin{equation}
S=\frac{1}{4\pi \alpha'}\int_\Sigma d^2\sigma \sqrt{-h}h^{ab}
\partial_a X^{\mu}\partial_b X^{\nu}
-\int_{\partial \Sigma} ds 
  \left( A_{\mu}\frac{\partial}{\partial s}X^{\mu}
         +\phi^{i}\frac{\partial}{\partial \sigma}X^i 
  \right)   
\label{osigma}
\end{equation}
for bosonic case where $X^i$s denote the Dirichlet directions.  If we
consider supersymmetric case, we should consider a suitable
supersymmetric generalization of (\ref{osigma}). For the flat
background, such model were considered in \cite{callan} in the RNS
formalism.  Even though the detailed form of the action is not known
for the plane wave background, we expect that the one dimensional
boundary theory defined above should have different form for the
boundary located away from the origin from that at the origin, since
the number of supersymmetries are different. Part of such differences
can be captured by the Dirac-Born-Infeld action, which can be derived
by the condition of the vanishing beta function of (\ref{osigma}) or
its suitable supersymmetric generalization. Currently this issue is on
the investigation \cite{hyun00}.
  
So far we work out the kappa variation up to the quartic terms in
$\theta$ coordinates. Thus it is interesting to see if the results
obtained above are persistent at the higher orders, which we suspect
so.  Especially for the half BPS branes where the analysis in the
light cone gauge is available, the higher terms should not modify the
analysis at the quartic order.  For the type IIB case, there are some
hand waving argument that the quartic results will go through the
higher orders \cite{bai038}.  It will be interesting to see if we can
find similar argument in the type IIA theory.

\section{D-particle}
Now we consider the possible constraints on the supersymmetry of the
open string where the D-particle boundary condition is given. This case
cannot be covered by the lightcone analysis.  As a first attempt, we
take $P=\Gamma^+$ which means the D-particle whose worldline lies
along $x^+$. Same thing happens for $P=\Gamma^-$.  Then even the
boundary contribution (\ref{b1}) does not vanish.  So we consider the
boundary condition $\theta^2 = P \theta^1$ with
\begin{equation}
P = \frac{1}{\gamma} ( \Gamma^+ + \gamma \Gamma^- ) ~,
\end{equation}
where $\gamma$ is a real constant.  With this boundary condition, one
can easily see that Eq. (\ref{b1}) vanishes.

Let's turn to the boundary contribution, Eq. (\ref{b2}).  First look at
the term with the structure of $\bar{\theta} \Gamma_{rs} \delta
\theta$, the third line in (\ref{b2}).  It does not vanish when $r \in
N$ and $s \in D$, and is proportional to
\begin{equation}
(\bar{\theta}^1 \Gamma_{rs} P \delta \theta^1 )
[ \bar{\theta}^1 \Gamma^s \Gamma^4 ( \Gamma^+ \theta^1_-
 + \gamma \Gamma^- \theta^1_+ ) ] ~,
\end{equation}
where 
\begin{equation}
\theta^1_\pm = h_\pm \theta^1 ~.
\end{equation}
We should impose additional boundary conditions as
\begin{equation}
\Gamma^\pm \theta^1_\mp \Big|_{\partial \Sigma} = 0 ~.
\label{d0b1}
\end{equation}

The terms in the first line of (\ref{b2}) vanishes basically because
of the anti-symmetric property between the indices $r$ and $s$, and
$r,s \in N$.  (If $r \in D$ or $s \in D$, it automatically vanishes.)

The terms in the second line of (\ref{b2}) do not vanish.  For
example, for $s \in N$, that is, $\Gamma^s = P$, they are proportional
to, with (\ref{d0b1}),
\begin{equation}
(\bar{\theta}^1_+ \Gamma^+ \delta \theta^1_+
+ \gamma \bar{\theta}^1_- \Gamma^- \delta \theta^1_- )
(\theta^1_+ \Gamma^4 \Gamma^- \theta^1_-
 - \theta^1_- \Gamma^4 \Gamma^+ \theta^1_+ ) ~.
\end{equation}
To eliminate this contribution, we should further require boundary
condition as
\begin{equation}
\Gamma^\pm \theta^1_\pm \Big|_{\partial \Sigma} = 0 ~.
\label{d0b2}
\end{equation}

All other terms with each boundary conditions for indices vanish.  For
example, the first term vanishes basically because of the
anti-symmetric property between the indices $r$ and $s$, and $r,s \in
N$.  (If $r \in D$ or $s \in D$, it automatically vanishes.)

We see that D-particle is not supersymmetric.

\section*{Acknowledgments}
The work of S.H. was supported in part by grant No.
R01-2000-000-00021-0 from the Basic Research Program of the Korea
Science and Engineering Foundation. The work of J.P. was supported by
Korea Research Foundation (KRF) Grant KRF-2002-070-C00022 and by
POSTECH BSRI research fund 1RB0210601. The work of H.S. was supported
by KRF Grant KRF-2001-015-DP0082.


\begin{thebibliography}{99}

\bibitem{met044} R. R. Metsaev, ``Type IIB Green-Schwarz superstring
  in plane wave Ramond-Ramond background,'' Nucl. Phys.  {\bf B625}
  (2002) 70, hep-th/0112044.

\bibitem{ber021} D. Berenstein J. Maldacena and H. Nastase, ``Strings
  in flat space and pp waves from ${\cal N} = 4$ Super Yang Mills,''
  JHEP {\bf 0204} (2002) 013, hep-th/0202021.

\bibitem{bla242} M. Blau, J. Figueroa-O'Farrill, C. Hull and G.
  Papadopoulus, ``A new maximally supersymmetric background of IIB
  superstring theory,'' JHEP {\bf 0201} (2001) 047, hep-th/0110242.

\bibitem{bla081} M. Blau, J. Figueroa-O'Farrill, C. Hull and G.
  Papadopoulus, ``Penrose limits and maximal supersymmetry,'' Class.
  Quant. Grav. {\bf 19} (2002) L87, hep-th/0201081.

\bibitem{sug} Y. Kiem, Y. Kim, S. Lee, and J. Park,
``pp wave/Yang-Mills correspondence: an explicit check,'' 
Nucl. Phys. {\bf B642} (2002) 389, hep-th/0205279.

\bibitem{sug1} P. Lee, S. Moriyama, and J. Park, ``A note on cubic
interactions in pp wave light cone string field theory,''
hep-th/0209011; P. Lee, S. Moriyama, and J. Park, ``Cubic interactions
in pp-wave light-cone string field theory,'' Phys. Rev.  {\bf D66}
(2002) 085021, hep-th/0206065.
\bibitem{parity} 
C.-S. Chu, V.V. Khoze, M. Petrini, R. Russo, and A. Tanzini,
``A note on string interaction on the pp wave
background,'' hep-th/0208148; 
C.-S. Chu, M. Petrini, R. Russo, and A. Tanzini, ``String
interactions and discrete symmetries of the pp-wave background,''
hep-th/0211188.

\bibitem{str} M. Spradlin and A. Volovich, ``Superstring interactions
in a pp wave background,'' Phys. Rev. {\bf D66} (2002) 086004,
hep-th/0204146; M. Spradlin and A. Volovich, ``Superstring
interactions in a pp wave background 2,'' hep-th/0206073.

\bibitem{ym} C. Kristjansen, J. Plefka, G.W. Semenoff, and
M. Staudacher, ``A new double scaling limit of N=4 super Yang-Mills
theory and pp wave strings,'' Nucl. Phys. {\bf B643} (2002) 3,
hep-th/0205033; N.R. Constable, D.Z. Freedman, M. Headrick,
S. Minwalla, L. Motl, A. Postnikov, and W. Skiba, ``pp wave string
interactions from perturbative Yang-Mills theory,'' JHEP {\bf 0207}
(2002) 017, hep-th/0205089; I.R. Klebanov, M. Spradlin, and
A. Volovich, ``New effects in gauge theory from pp wave
superstrings,'' Phys. Lett. {\bf B548} (2002) 111, hep-th/0206221;
J. Gomis, S. Moriyama, and J. Park, ``SYM description of SFT
Hamiltonian in a pp wave background,'' hep-th/0210153;

\bibitem{mix} U. Gursoy, ``Vector operators in the BMN
correspondence,'' hep-th/0208041; N. Beisert, C. Kristjansen,
J. Plefka, G.W. Semenoff, and M.  Staudacher, ``BMN correlators and
operator mixing in N=4 superYang-Mills theory,'' hep-th/0208178;
D.J. Gross, A. Mikhailovi, and R. Roiban, ``A calculation of the plane
wave string Hamiltonian from N=4 superYang-Mills theory,''
hep-th/0208231; N.R. Constable, D.Z. Freedman, M. Headrick, and
S. Minwalla, ``Operator mixing and the BMN correspondence,'' JHEP {\bf
0210} (2002) 068, hep-th/0209002; R.A. Janik, ``BMN operators and
string field theory,'' hep-th/0209263.

\bibitem{bit} H. Verlinde, ``Bits, Matrices and 1/N,'' hep-th/0206059;
J.-G. Zhou, ``pp wave string interactions from string bit model,''
hep-th/0208232; D. Vaman and H. Verlinde, ``Bit strings from N=4 gauge
theory,'' hep-th/0209215; J. Pearson, M. Spradlin, D. Vaman,
H. Verlinde, and A. Volovich, ``Tracing the string: BMN correspondence
at finite $J^2 /N$,'' hep-th/0210102.



\bibitem{das185} K. Dasgupta, M. M. Sheikh-Jabbari, M. Van Raamsdonk,
  ``Matrix Perturbation Theory For M-Theory On a PP-Wave,'' JHEP {\bf
    0205} (2002) 056, hep-th/0205185.

\bibitem{hyu074} S. Hyun and H. Shin, ``N=(4,4) Type IIA String Theory
  on PP-Wave Background,'' JHEP {\bf 0210} (2002) 070, hep-th/0208074.
  
\bibitem{sug029} K. Sugiyama and K. Yoshida, ``Type IIA String and
  Matrix String on PP-wave,'' Nucl. Phys. {\bf B644} (2002) 128,
  hep-th/0208029.
  
\bibitem{hyu158} S. Hyun and H. Shin, ``Solvable N=(4,4) Type IIA
  String Theory in Plane-Wave Background and D-Branes,''
  hep-th/0210158.

\bibitem{hyu090} S. Hyun and H. Shin, ``Branes from Matrix Theory in
  PP-Wave Background,'' Phys. Lett. {\bf B543} (2002) 115,
  hep-th/0206090.

\bibitem{par161} J.-H. Park, ``Supersymmetric objects in the M-theory 
on a pp-wave,'' hep-th/0208161.

\bibitem{lam031} N. D. Lambert and P. C. West, ``D-Branes in the
  Green-Schwarz Formalism,'' Phys. Lett. {\bf B459} (1999) 515,
  hep-th/9905031.
  
\bibitem{bai038} P. Bain, K. Peeters and M. Zamaklar, ``D-branes in a
  plane wave from covariant open strings,'' hep-th/0208038.

\bibitem{dab231} A. Dabholkar and S. Parvizi, ``Dp Branes in PP-wave
  Background,'' Nucl. Phys. {\bf B641} (2002) 223, hep-th/0203231.
  
\bibitem{bil028} M. Billo and I. Pesando, ``Boundary states for GS
  superstrings in an Hpp wave background,'' Phys.\ Lett.\ B {\bf 536}
  (2002) 121, hep-th/0203028.

\bibitem{ske054} K. Skenderis and M. Taylor, ``Branes in AdS and
  pp-wave spacetimes,'' JHEP {\bf 0206} (2002) 025, hep-th/0204054.

\bibitem{gab122} M. R. Gaberdiel and M. B. Green, ``The D-instanton
  and other supersymmetric D-branes in IIB plane-wave string theory,''
  hep-th/0211122.

\bibitem{dpp} J. Morales, ``String theory on Dp-plane waves,'' hep-th/0210229.

\bibitem{cvetic} M. Cvetic, H. Lu, C. N. Pope and K. S. Stelle, ``Linearly-realized 
worldsheet supersymmetry in pp-wave background,'' hep-th/0209193. 


\bibitem{ske184} K. Skenderis and M. Taylor, ``Open strings in the
  plane wave background I: Quantization and symmetries,''
  hep-th/0211011; ``Open strings in the plane wave background II:
  Superalgebras and Spectra,'' hep-th/0212184.

\bibitem{dew209} B. de Wit, K. Peeters, J. Plefka and A. Sevrin, ``The
  M-theory two-brane in $AdS_4 \times S^7$ and $AdS_7 \times S^4$,''
  Phys. Lett. {\bf B443} (1998) 153, hep-th/9808052; P. Claus, ``Super
  M-brane actions in $AdS_4 \times S^7$ and $AdS_7 \times S^4$,''
  Phys. Rev. {\bf D59} (1999) 066003, hep-th/9809045.

\bibitem{pen271} R. Penrose, ``Any space-time has a plane wave
  limit,'' in {\it Differential Geometry and Gravity}, Reidel,
  Dordrecht 1976, pp. 271.

\bibitem{duf70} M. J. Duff, P. S. Howe, T. Inami and K. S. Stelle,
  ``Superstrings in $D=10$ from supermembranes in $D=11$,''
  Phys.~Lett.  {\bf B191} (1987) 70.
  
\bibitem{callan} C. G. Callan, C. Lovelace, C. R. Nappi and S. A.
  Yost, ``Loop corrections to superstring equations of motion,'' Nucl.
  Phys. {\bf B308} (1988) 221.

\bibitem{hyun00} S. Hyun, J. Park and H. Shin, in progress.

\end{thebibliography}
\end{document}